\documentclass[lettersize,journal,twoside]{IEEEtran}
\usepackage{amsmath,amsfonts,yhmath}
\usepackage{algorithmic}
\usepackage{algorithm}
\usepackage{array}
\usepackage[caption=false,font=footnotesize,labelfont=rm,textfont=sf]{subfig}
\usepackage{textcomp}
\usepackage{stfloats}
\usepackage{url}
\usepackage{verbatim}
\usepackage{graphicx}
\usepackage{cite}
\usepackage{multirow}

\begin{document}

\title{Fighting Fire with Fire: Channel-Independent RF Fingerprinting via the Ratio of Linear to Logarithmic Differential Spectrum}
\author{Tianshu Chen,~\IEEEmembership{Graduate Student Member,~IEEE,}
Aiqun Hu,~\IEEEmembership{Senior Member,~IEEE,}
Shiqi Zhang

        % <-this % stops a space
\thanks{Manuscript received XX; revised XX. \textit{(Corresponding author: Aiqun Hu.)}}
\thanks{Tianshu Chen, Aiqun Hu and Shiqi Zhang are with the National Mobile Communications Research Laboratory, Southeast University, Nanjing 210096, China (e-mail: iamtianshu@seu.edu.cn; aqhu@seu.edu.cn; zhangshq@seu.edu.cn).}
\thanks{This work has been submitted to the IEEE for possible publication. Copyright may be transferred without notice, after which this version may no longer be accessible.}
}

\markboth{IEEE Wireless Communications Letters,~Vol.~XX, No.~XX, XX~2024}{\protect\parbox{0.95\textwidth}{Chen \MakeLowercase{\textit{et al.}}: Fighting Fire with Fire: Channel-Independent RF Fingerprinting via the Ratio of Linear to Logarithmic Differential Spectrum}}

\IEEEpubid{0000--0000/00\$00.00~\copyright~2021 IEEE}

\maketitle

\begin{abstract}
  Eliminating the influence of temporally varying channel components on the radio frequency fingerprint (RFF) extraction has been an enduring and challenging issue. To overcome this problem, we propose a channel-independent RFF extraction method inspired by the idea of `fighting fire with fire'. Specifically, we derive the linear differential spectrum and the logarithmic differential spectrum of the channel frequency responses (CFRs) from the received signals at different times, and then calculate the ratio of the two spectrums. It is found that the division operation effectively counteracts the channel effects, while simultaneously preserving the integrity of the RFFs. Our experiments on LTE-V2X, LoRa and Wi-Fi devices show that the proposed method achieves an average identification accuracy exceeding 95\% across various environments.
\end{abstract}


%\vspace{-0.5 cm}

\section{Introduction}

\IEEEPARstart{W}{ith} the advancement of mobile communication technologies, particularly the maturation of 5G and emerging 6G technologies\cite{2023Mitev}, our society has entered an era characterized by intelligent Internet of Things (IoT) connectivity\cite{2023Ferrag}. Nevertheless, the openness of wireless networks, coupled with existing vulnerabilities in authentication mechanisms, exposes the communication systems to various security threats such as identity spoofing, tampering, session hijacking\cite{2024JingWT}. The conventional authentication technologies, while effective, usually rely on cryptography algorithms or cloud-based encryption\cite{2024Javadpour}, which demand substantial computational resources\cite{2024LanXL} and remain susceptible to brute-force attacks from quantum computers\cite{2020Luo}.

In recent years, the radio frequency fingerprint (RFF) technology has emerged as a promising physical layer authentication mechanism that can be combined with cryptographic techniques to further improve security\cite{2022Jagannath}. The RFF is an inherent hardware characteristic arising from manufacturing imperfections in the analog circuitry, which will naturally manifest itself during signal transmission\cite{2020Soltanieh}. The unique hardware features are stable and unforgeable, thus can be treated as reliable fingerprints of the devices\cite{2024YaoZS}. Furthermore, the RFF identification system only needs to be deployed at the receiver side, without consuming computational resources at the transmitter side or altering the original communication system\cite{2024ShenGX}. This allows for rapid `receipt-as-authentication', promptly delivering identification results upon signal reception.

However, in mobile communication systems, wireless terminals often operate in time-varying channel environments, which introduces some interference in extracting stable RFF features from the received signals. The channel impulse response (CIR) and the transmitted signal are typically modeled as a convolution relationship, hence complicating the removal of channel components while preserving RFF features.

The current channel-resilient RFF extraction methods can be broadly categorized into three groups. The first group utilizes frequency spectrum division\cite{2022ShenGX,2024PengLN} or differential operations\cite{2024PengLN2} to eliminate the channel information based on the assumption that the channel frequency responses (CFRs) of adjacent signal segments are approximately identical. However, this assumption is difficult to maintain in highly dynamic mobile scenarios. Furthermore, the convolutional fingerprints will also be inadvertently removed during the division or differential operations, leading to a partial loss of RFF information. The second group employs channel estimation\cite{2021Fadul,2022ChenTS2} or filter\cite{2023MengXQ} to isolate multipath channel effects. However, the channel and RFF elements are often intertwined and cannot be completely separated by channel equalization or filtering techniques. The third group includes methods based on deep learning\cite{2023XieRJ}, but the unpredictability of wireless channel environments makes it challenging to ensure the training sufficiency, potentially degrading model generalization when encountering new conditions. In general, while existing methods have made significant progress, there remains substantial room for enhancing the identification accuracy.

\IEEEpubidadjcol
To overcome the aforementioned limitations, we propose a lightweight method for extracting channel-independent RFF features in this letter, which is based on the idea of `fighting fire with fire'. Specifically, the channel component is eliminated through a division operation between the difference of linear spectrum and the difference of logarithmic spectrum of the CFRs from the received signals at different times. By leveraging the variability of the wireless channel across multiple temporal instances, the time-varying channel components can be reversely counteracted, ultimately yielding pure and stable RFFs. Our experiments use devices based on three different communication technologies, i.e. LTE-vehicle-to-everything (LTE-V2X), LoRa, and Wi-Fi, to validate the identification performance of the proposed method. In cross-scenario experiments conducted at speeds not exceeding 30 km/h, the average accuracy exceeds 95\% when training with wired-connected data and testing with data from different wireless environments.

The rest of this letter is organized as follows. Section II introduces the system framework. Section III elaborates on the proposed channel-independent RFF extraction methodology based on the idea of `fighting fire with fire'. Section IV evaluates the performance of the proposed RFF extraction method through experiments. Finally, Section V concludes this letter.

\begin{figure}[!t]
  \centering
  \includegraphics[width=7cm]{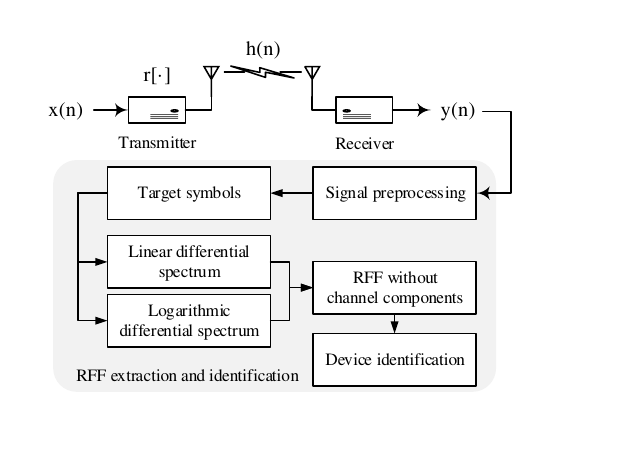}
  \caption{Diagram of the proposed RFF extraction and identification system.}
  \label{fig1}
\end{figure}

\section{System Overview}

Fig.\ref{fig1} demonstrates the framework of the RFF extraction and identification system, from signal transmission to reception, followed by feature extraction and device identification. Initially, the baseband signal $x(n)$ undergoes modulation and up-conversion into an RF signal for transmission. The RFF of the transmitter caused by the hardware impairments will result in distortion of the transmitted signal, which is recorded as $r[x(n)]$. Then the distorted signal will propagate through the wireless channel before reaching the receiver, where the time-domain channel impulse response is recorded as $h(n)$. Afterwards, the received signal is down-converted from the RF band to the baseband and sampled as $y(n)$, which is mathematically expressed as
\begin{equation}
  y(n)=r[x(n)]\ast h(n)+z(n), 0 \leq n \leq N-1,
  \label{eq21}
\end{equation}
in which $z(n)$ represents the additive white Gaussian noise (AWGN), $N$ denotes the number of sampling points, and $\ast$ signifies the convolution operation.

Upon obtaining $y(n)$, the signal will be sent to RFF extraction and identification system, involving preprocessing and channel components elimination. Finally, after extracting the RFF features, device identification is performed. The detailed methodology for RFF extraction will be elaborated in the next section.

\section{Proposed RFF Extraction Method}

In this section, we present our proposed RFF extraction method based on the strategy of `eliminating the channel components with changing CFR', which is inspired by the idea of `fighting fire with fire'. Specifically, the method eliminates the impact of channel effects on RFFs by exploiting the variances in both the linear and logarithmic spectrums from the CFRs of the received signals at distinct times. The proposed method comprises three stages: signal preprocessing, target symbols selection, and CFR elimination.

\subsection{Signal Preprocessing}
In order to eliminate the interference from irrelevant factors and ensure stable RFF extraction, the signal preprocessing procedure includes time synchronization, carrier frequency offset (CFO) compensation and energy normalization. Time synchronization is used to determine the starting point of each frame. Inaccurate synchronization will cause temporal misalignment in the subsequent selection of target symbols. CFO compensation corrects for frequency drift caused by temperature variations and other environmental factors, which can otherwise degrade the stability of RFFs. Energy normalization is carried out to eliminate the influence of the received signal strength indicator on RFF identification. The detailed descriptions can be found in our prior work \cite{2022ChenTS2}.

\subsection{Selection of Target Symbols}

To capitalize on the temporal channel variability to eliminate channel effects, it is essential to select two received symbols that experience different channel conditions. Consequently, the time interval $\tau$ between the two symbols should satisfy $\tau>\Delta $, where $\Delta$ is the minimum incoherent time of the channel.

The preprocessed time-domain received signal is transformed into the frequency domain by performing discrete Fourier transform (DFT) represented as
\begin{equation}
  Y(k)=\mathcal{R} [X(k)]H(k), 0 \leq k \leq N-1,
  \label{eq22}
\end{equation}
where $\mathcal{R} [X(k)]$ denotes the frequency domain expression of $r[x(n)]$, and $H(k)$ denotes the channel frequency response (CFR). Correspondingly, the frequency-domain received signals of the two selected symbols are denoted as $Y_1 (k)$ and $Y_2 (k)$, referring to the preprocessed signal illustrated in \eqref{eq22}. To facilitate the extraction of stable RFFs, we select two symbols whose local signals are both $X (k)$, e.g. identical sequences in preamble, while ensuring that they experience dissimilar channel characteristics, denoted as $H_1 (k)$ and $H_2 (k)$.

\subsection{`Fighting Fire with Fire'}

Subsequently, a differential operation is performed on the linear spectrums and the logarithmic spectrums of the two received signals, which is respectively given by
\begin{equation}
  D(k)=Y_1 (k)-Y_2 (k)=\mathcal{R} [X(k)][H_1 (k)-H_2 (k)],
  \label{eq31}
\end{equation}
and
\begin{equation}
  \begin{aligned}
  L(k)&=\ln Y_1 (k)-\ln Y_2 (k) =\ln \frac{\mathcal{R} [X(k)] H_1 (k)}{\mathcal{R} [X(k)] H_2 (k)} \\
  &=\ln H_1 (k)-\ln H_2 (k),
  \label{eq32}
\end{aligned}
\end{equation}
where $D(k)$ and $L(k)$ represent the linear differential spectrum and the logarithmic differential spectrum, respectively. By applying logarithmic differential operation, we aim to approximate the difference in the logarithm of CFR at different times to the difference in the CFR itself, which facilitates subsequent elimination of channel components in conjunction with the linear differential spectrum.

Then, the CFR $H(k)$ can be further written as
\begin{equation}
  H(k)=1+\Delta H(k).
  \label{eq33}
\end{equation}
In an ideal scenario, the frequency response corresponding to the unit impulse response is a constant value of 1. While in practical wireless communications, multipath propagation results in frequency-selective fading, causing fluctuations of the CFR, which is recorded as $\Delta H(k)$. According to the Taylor series expansion, when $\Delta H(k) \to 0$, $\ln [1+\Delta H(k)]$ can be expanded as
\begin{equation}
  \begin{aligned}
  \ln [1+\Delta H(k)]=\Delta H(k)-\frac{1}{2}[\Delta H(k)]^2+\frac{1}{3}[\Delta H(k)]^3\\
  -\cdots +\frac{(-1)^{n-1}}{n}[\Delta H(k)]^n+O([\Delta H(k)]^{n+1}).
  \label{eq34}
  \end{aligned}
\end{equation}

The error associated with approximating $\ln [1+\Delta H(k)]$ to $\Delta H(k)$ can be expressed as
\begin{equation}
  \begin{aligned}
  E(k)&=\Delta H(k)-\ln [1+\Delta H(k)]\\
  &=\frac{1}{2}[\Delta H(k)]^2-\frac{1}{3}[\Delta H(k)]^3+\cdots.
  \label{eq41}
  \end{aligned}
\end{equation}
For extremely small values of $\Delta H(k)$, the higher-order terms can be ignored, and therefore
\begin{equation}
  E(k)\approx \frac{1}{2}[\Delta H(k)]^2.
  \label{eq42}
\end{equation}
When the approximation error is much smaller than the amplitude of RFF frequency response, i.e.,
\begin{equation}
  |E(k)|\ll \left\lvert\frac{\mathcal{R} [X(k)]}{X(k)}\right\rvert -1,
  \label{eq43}
\end{equation}
the impact of this error on RFF recognition becomes negligible. Accordingly, we can approximate $\ln [1+\Delta H(k)]$ as $\Delta H(k)$ under conditions where the fluctuation range of CFR is small enough to satisfy \eqref{eq43}, in which $E(k)$ is substituted by \eqref{eq42}. In such circumstances, \eqref{eq32} can be rewritten as
\begin{equation}
  \begin{aligned}
  L(k)&=\ln \left[1+\Delta H_1(k)\right]-\ln \left[1+\Delta H_2(k)\right] \\
  &\approx \Delta H_1(k)-\Delta H_2(k)=H_1 (k)-H_2 (k).
  \label{eq35}
  \end{aligned}
\end{equation} 

Specifically, for broadband signals, to avoid the possibility that deep fading could lead to channel fluctuations failing to meet \eqref{eq43}, we divide the broadband channel into several narrow subbands and normalize the average amplitude of each narrowband channel to 1. Following this, we extract RFFs for each narrowband segment.

Afterwards, the result of the aforementioned linear differential operation is divided by the outcome of the logarithmic differential operation to obtain the initial RFF without channel components, which is calculated by
\begin{equation}
  F(k)=\frac{D(k)}{L(k)} =\frac{\mathcal{R} [X(k)][H_1 (k)-H_2 (k)]}{H_1 (k)-H_2 (k)}=\mathcal{R} [X(k)].
  \label{eq36}
\end{equation}
This equation effectively counteracts the channel difference term $H_1 (k)-H_2 (k)$ in the two differential spectrums via ratio operation, ensuring that the outcomes reflect only the intrinsic hardware characteristics of the transmitter.

Finally, to alleviate the impact of noise on the extracted RFF, we employ the repeated sequences in the preamble and average the RFF features corresponding to $M$ groups of target symbols that satisfy the conditions described in Section III-B, as expressed by
\begin{equation}
  \overline{F}(k)=\frac{1}{M} \sum_{m = 1}^{M} F_m(k),
  \label{eq37}
\end{equation}
where $F_m(k)$ denotes the initial RFF of the $m$-th group of target symbols. $\overline{F}(k)$ is the ultimate RFF expression without channel components and noise.

\section{Experiment Verification}

\subsection{Experiment Setup}

In the experiment, we choose devices based on three different communication technologies as transmitters, including 12 LTE-V2X modules (Morningcore CX7100), 10 LoRa modules (Heltec HTCC-AB02A), and 28 Wi-Fi routers (20 Mercury MW305R and 8 Dlink DWL-2000AP+A). A universal software radio peripheral (USRP) is utilized as the receiver. The bandwidths of the three types of signals are configured as 20 MHz, 500 kHz, and 20 MHz, while the carrier frequencies are set to 5.915 GHz, 433 MHz, and 2.472 GHz, respectively. Since the local signals of fixed sequences can be readily obtained, e.g. primary sidelink synchronization signal (PSSS), secondary sidelink synchronization signal (SSSS), and demodulation reference signal (DMRS) in LTE-V2X, linear chirps in LoRa, and long training symbol (LTS) in Wi-Fi, we choose these sequences as target symbols for RFF extraction.

\begin{table*}[!t]
  \caption{Classification Accuracy Under Different Experimental Environments and Different Classification Algorithms\label{accuracy}}
  \centering
  \begin{tabular}{c|c|c c c|c c c}
  \hline
  \multirow{2}*{Experimental devices} &\multirow{2}*{Training dataset} & \multicolumn{3}{c|}{Test dataset} & \multicolumn{3}{c}{Accuracy (\%)}\\ 
  %\cline{2-8}
  ~ & ~ & Location & Channel & Speed & Random forest & XGBoost & LSTM-MLP\\
  \hline
  \multirow{7}*{12 LTE-V2X modules} & \multirow{7}*{Wired connection} & Indoor & Wireless LOS & 1-5 km/h & 97.17 & 97.42 & 88.25 \\
  ~ & ~ & Corridor & Wireless NLOS & 1-5 km/h & 96.83 & 96.17 & 89.25 \\
  ~ & ~ & Outdoor & Wireless LOS+NLOS & 10-30 km/h & 93.50 & 88.42 & 79.92 \\
  ~ & ~ & Outdoor & Simulated LTE ETU & 30 km/h & 93.17 & 89.67 & 76.42 \\
  ~ & ~ & Outdoor & Simulated LTE ETU & 60 km/h & 91.42 & 86.83 & 71.83 \\
  ~ & ~ & Outdoor & Simulated LTE ETU & 90 km/h & 89.83 & 85.08 & 70.33 \\
  ~ & ~ & Outdoor & Simulated LTE ETU & 120 km/h & 88.75 & 81.58 & 68.25 \\
  \hline
  \multirow{4}*{10 LoRa modules} & \multirow{4}*{Wired connection} & Indoor & Wireless LOS & 0 km/h & 98.90 & 98.20 & 99.00 \\
  ~ & ~ & Indoor & Wireless LOS & 1-5 km/h & 98.40 & 96.70 & 98.60 \\
  ~ & ~ & Corridor & Wireless NLOS & 0 km/h & 98.80 & 98.20 & 98.70 \\
  ~ & ~ & Corridor & Wireless NLOS & 1-5 km/h & 98.60 & 97.60 & 98.10 \\
  \hline
  \multirow{4}*{28 Wi-Fi routers} & \multirow{4}*{Wired connection} & Indoor & Wireless LOS & 0 km/h & 98.77 & 98.55 & 97.39 \\
  ~ & ~ & Indoor & Wireless LOS & 1-5 km/h & 93.25 & 89.21 & 88.43 \\
  ~ & ~ & Corridor & Wireless NLOS & 0 km/h & 96.61 & 86.64 & 82.79 \\
  ~ & ~ & Corridor & Wireless NLOS & 1-5 km/h & 92.29 & 80.32 & 81.14 \\
  \hline
  \end{tabular}
\end{table*}

The experimental environments outlined in Table \ref{accuracy} are elaborated as follows. First, the devices are directly connected to the USRP via a wired manner as training data, thereby avoiding the classifier learning channel characteristics. Then we collect the signals in the wireless environments as test data, where the USRP is placed in a fixed position and the experimental devices are placed on a trolley that traverses random paths. The experiments are conducted indoors, in corridors and outdoors, covering static and mobile environments, as well as line-of-sight (LOS) and non-line-of-sight (NLOS) scenarios. Finally, for LTE-V2X modules, we pass the signals collected in a static outdoor environment through a simulated LTE Extended Typical Urban (ETU) multipath fading channel model\cite{3gpp36101}. Additionally, a Doppler frequency shift ranging from 164 to 655 Hz is added, which corresponds to a vehicle speed of 30 to 120 km/h. This process enables us to obtain the data in high-speed environments. The average signal-to-noise ratio (SNR) of the collected signals approximates 20 dB.

\subsection{Selection of Valid Experimental Data}

In each experimental environment, the collected signal frames are sorted chronologically and divided into two equal groups. The $i$-th frame from each group is then paired for analysis to ensure a sufficient time interval between them. Next, two symbols with identical local signals, respectively from the paired frames, are chosen to calculate the initial RFF. For frames containing multiple types of symbols, we first independently average the initial RFFs for each symbol type using \eqref{eq37}, and then concatenate them to construct the ultimate RFF expression for device identification.

\begin{figure}[!t]
  \centering
  \includegraphics[width=8.5cm]{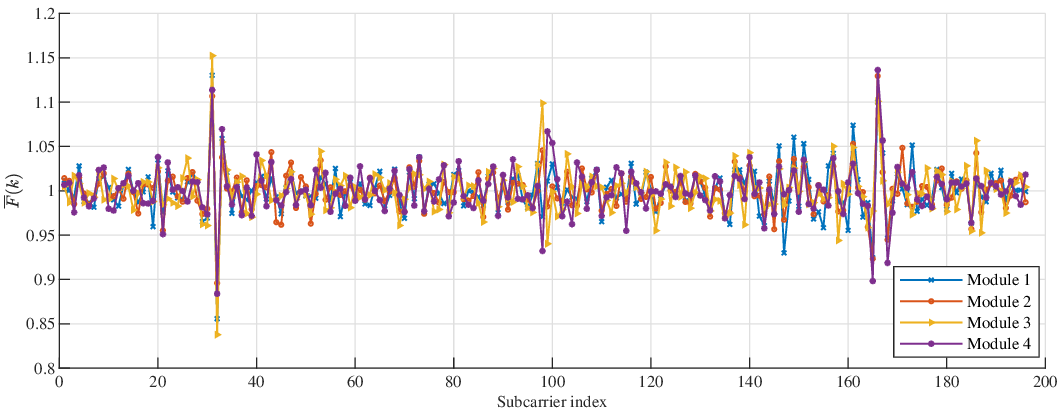}
  \caption{RFFs of 4 different LTE-V2X modules.}
  \label{RFFdif}
\end{figure}

To meet the condition specified in \eqref{eq43}, we use CFRs estimated from wired connections to obtain the amplitude of the RFF frequency response and subsequently determine the domain of $\Delta H(k)$, as the channel effect is negligible compared to the RFF characteristics under this condition. Fig. \ref{RFFdif} visually displays the ultimate RFF waveforms $\overline{F}(k)$ of partial LTE-V2X modules. The RFF amplitude of the devices in our experiment ranges from 0.85 and 1.15, i.e. fluctuating within $\pm$0.15. Accordingly, the approximation error should remain one order of magnitude smaller than RFF amplitude, hence we constrain $\Delta H(k)$ to be within 0.173. Symbols with channel fluctuations exceeding this threshold are excluded.

\subsection{Experiment results}

\begin{figure}[!t]
  \centering
  \subfloat[]{\includegraphics[width=4.2cm]{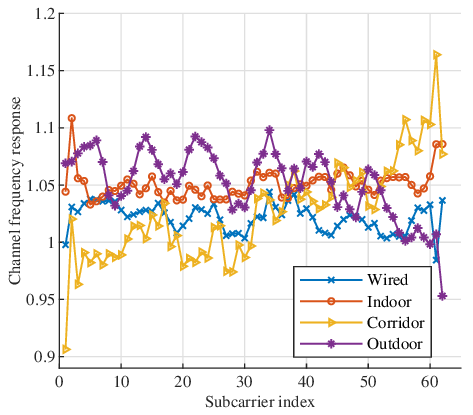}
  \label{CFR}}
  \hfil
  \subfloat[]{\includegraphics[width=4.2cm]{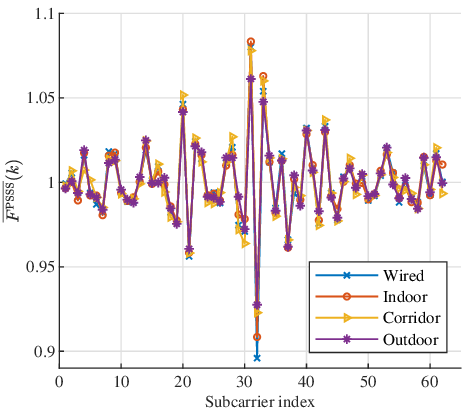}
  \label{RFF}}
  
  \caption{The CFR and RFF extracted from the PSSS symbols of the same LTE-V2X module in different experimental environments. (a) CFR waveforms. (b) RFF waveforms.}
  \label{CFRRFF}
\end{figure}

First, we take an LTE-V2X module as an example and compare the CFR and RFF waveforms of the measured data in different experimental environments in Fig. \ref{CFRRFF}. The differences of the CFRs depicted in Fig. \ref{CFRRFF}\subref{CFR} arise from the variability in channel fading across different scenarios, while the corresponding RFF features exhibit a high degree of consistency, as clearly shown in Fig. \ref{CFRRFF}\subref{RFF}. This indicates that our proposed method effectively eliminates the impact of channel effects on RFF feature extraction. On the other hand, the RFF features derived from different modules show significant differentiation, as illustrated in Fig. \ref{RFFdif}, which confirms the capability to distinguish different devices.

We then perform classification experiments on the three types of devices. For each training or test dataset, as presented in Table \ref{accuracy}, 100 groups of valid RFF feature vectors are selected for each module. The complex feature vectors $\overline{F}(k)$ are divided into real and imaginary parts before being fed into the classifier. The classification accuracy under different experimental environments and different classification algorithms is summarized in Table \ref{accuracy}. The training times of the three classification models are 1.738 s, 8.615 s, and 2.663 s, respectively. The results reveal that the random forest algorithm outperforms both the extreme gradient boosting (XGBoost) algorithm and the long short-term memory-multilayer perceptron (LSTM-MLP) algorithm in terms of higher accuracy and faster training speed. In cross-scenarios tests at speeds up to 30 km/h, we achieve average classification accuracies of 95.83\%, 98.68\%, and 95.23\% for the three types of devices when using random forest algorithm. Even in high-speed simulation experiments, the accuracy still remains above 88.75\%. 

\subsection{Performance Comparison Across SNRs \& Benchmarks}

\begin{figure}[!t]
  \centering
  \includegraphics[width=7.5cm]{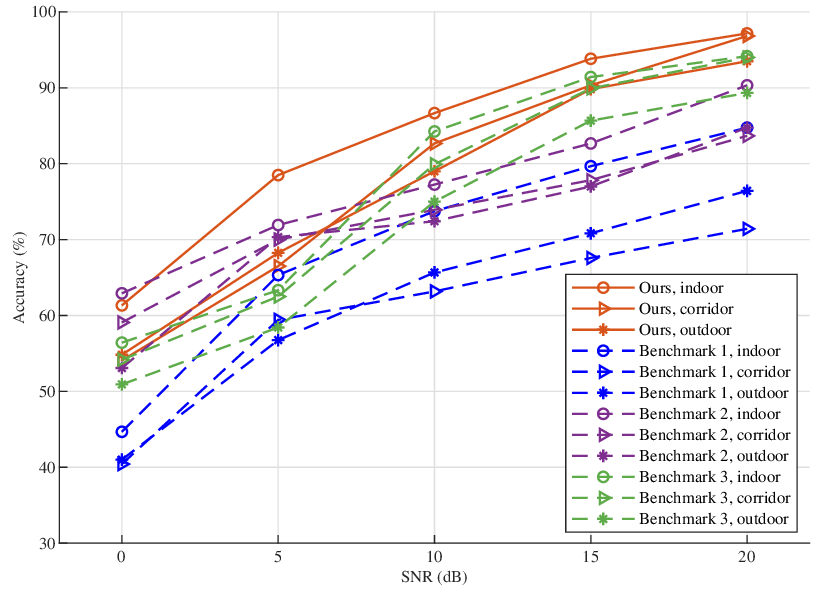}
  \caption{Identification accuracy versus SNR for our proposed method and benchmarks under cross-scenarios testing.}
  \label{snr}
\end{figure}

\begin{table}[!t]
  \caption{Computational Complexity Comparisons Between Our Proposed Method and Benchmarks\label{complexity}}
  \centering
  \begin{tabular}{c c c c}
  \hline
  \multirow{2}*{Method} & RFF representation  & Training time & Test time \\
  ~ & complexity & (s) & (ms)\\
  \hline
  Ours & $O(n)$ & 1.738 & 0.010 \\
  Benchmark 1 & $O(n\log n)$ & 2.830 & 0.011 \\
  Benchmark 2 & $O(n\log n)$ & 3.088 & 0.014 \\
  Benchmark 3 & $O(n^2)$ & 2.520 & 0.022 \\
  \hline
  \end{tabular}
\end{table}

In this subsection, we add AWGN to the origin time-domain received signals of 12 LTE-V2X modules to investigate the identification accuracy under different SNRs. The SNR ranges from 0 dB to 20 dB. Besides, we compare our method against three up-to-date RFF extraction techniques for LTE devices: hybrid feature matrix construction\cite{2024PengLN}, channel estimation based methodology\cite{2022ChenTS2} and temporal correlation-based scheme\cite{2024QiXY}. The identification accuracy versus SNR for our proposed method and the benchmarks is depicted in Fig. \ref{snr}. Our method outperforms all benchmarks, especially under high SNR conditions. Specifically, at 20 dB SNR, our method achieves average accuracy improvements of 18.30\%, 9.58\%, and 3.33\% over the three benchmark methods, respectively. When the SNR drops to 15 dB and 10 dB, the average accuracy of our method decreases by 4.50\% and 13.05\%, but still exceeds that of the benchmarks.

Furthermore, we compare the computational complexity between our method and the benchmarks, as summarized in Table \ref{complexity}. The results indicate that our method demonstrates the lowest computational complexity for RFF feature extraction. Additionally, both the training time of the classifier and the test time are also the shortest among the evaluated methods.

\section{Conclusion}
In this letter, a channel-independent RFF extraction method is proposed based on the idea of `fighting fire with fire'. Focusing on eliminating the impact of channel components on RFFs, we leverage the variability of the wireless channel across different temporal instances and calculate the ratio of their linear and logarithmic differential spectrums. Finally, we successfully obtain highly distinguishable and stable RFF features. As verified via experiments on LTE-V2X, LoRa and Wi-Fi devices, our method displays robust performance under challenging mobile environments and cross-scenarios testing.

\bibliographystyle{IEEEtran}
\bibliography{IEEEabrv,references}

\vfill

\end{document}